\documentclass{WileyMSP-template}
\usepackage{cite}
\usepackage{array}
\usepackage{microtype}
\usepackage{float}
\usepackage{ragged2e}
\begin{document}

\pagestyle{fancy}
\rhead{\includegraphics[width=2.5cm]{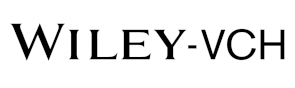}}
	
\title{{\Large Thin-shell wormhole under non-commutative geometry inspired Einstein-Gauss-Bonnet  gravity.}}

\maketitle
	
\author{N. Rahman}
\author{M. Kalam}
\author{A. Das}
\author{S. Islam}
\author{F. Rahaman}
\author{M. Murshid}


\begin{affiliations}
	N. Rahman\\
	Department of Physics, Aliah University, \\
	IIA/27, New Town, Kolkata 700160, India\\
	Email: rahmannilofar@gmail.com\\
	
	Dr. M. Kalam\\
	Department of Physics, Aliah University, \\
	IIA/27, New Town, Kolkata 700160, India\\
	Email: mehedikalam@yahoo.co.in\\
	
	Dr. A. Das\\
	Department of Physics, Ashoknagar Vidyasagar Bani Bhaban \\
	High School (H.S.), North 24 Parganas, West Bengal - 743222, India\\
	Email: amdphy@gmail.com\\
	
	Dr. S. Islam\\
	Department of Mathematics, Amity University Kolkata, Kolkata - 700135, India\\
	Email: sayeedul.jumath@gmail.com\\
	
	Dr. S. Islam\\
	Department of Mathematics, Sister Nivedita University, Kolkata - 700156, India\\
	Email: sayeedul.jumath@gmail.com\\
	
	Dr. F. Rahaman\\
	Department of Mathematics, Jadavpur University, Kolkata - 700032, India\\
	Email: rahaman@associates.iucaa.in\\
	
	M. Murshid\\
	Department of Physics, Aliah University, \\
	IIA/27, New Town, Kolkata 700160, India\\
	Email: masum.murshid@wbscte.ac.in\\
	
\end{affiliations}

\date{\today}	


\justify

\begin{abstract} 
	Einstein-Gauss-Bonnet gravity is a generalization of the general relativity to higher dimensions in which the ﬁrst and second-order terms corresponds to general relativity and Einstein-Gauss-Bonnet gravity respectively.
	We construct a new class of ﬁve-dimensional (5D)  thin-shell wormholes by the `Cut-Paste' technique  from black holes in Einstein-Gauss-Bonnet  gravity inspired  by non-commutative geometry starting with a static spherically symmetric, Gaussian mass distribution as a source and for this structural form of the thin shell wormhole we have explored several salient features of the solution, viz., pressure-density profile, equation of state, the nature of wormhole, total amount of exotic matter content at the shell. We have also analyzed the linearized stability of the constructed wormhole. From our study we can assert that our model is found to be plausible with reference to the other model of thin-shell wormhole available in literature.

\end{abstract}

\section{Intoduction}

In 1935, Albert Einstein and Nathan Rosen \cite{Einstein1935} tried to produce a ﬁeld theory for electrons. Considering a spherically symmetric mass distribution which had already been used for black holes, Einstein and Rosen implemented a co-ordinate transformation to eliminate the region comprising the curvature singularity. But, the solution demonstrates the existence of bridges through space-time. This bridge can act as a tunnel through which one can travel in space-time to another part of the universe and it  is known as Einstein-Rosen bridge. It is also known as Schwarzschild wormhole. Theoretically, this bridge is a shortcut path that minimizes the travel time as well as distance. However, in 1962, John Archibald Wheeler and Robert W. Fuller \cite{Wheeler1962} argued that Einstein and Rosen bridge structure is unstable if it connects two parts of the same universe. Therefore, one can argue that though Schwarzschild wormhole exists theoretically, but it is not traversable in both directions. In 1988, Michael S. Morris and Kip S. Thorne \cite{Morris1988} first ever proposed a wormhole solution of Einstein field equations which was found to be traversable. The solution represents a hypothetical shortcuts between two regions (either of the same Universe or of two separate Universes) connected by a throat. The throat of the wormholes can be deﬁned as a two dimensional hypersurface of minimal area and for the existance of such an wormhole the  null energy condition has to be violated. Hence, all traversable wormholes require exotic matter which can be held responsible for the violation of  the null energy condition. The requirements of exotic matter for the existence of a wormhole can be made inﬁnitesimally small by a suitable choice of the geometry \cite{Visser2003}. M. Visser \cite{Visser1989} has also proposed another way to minimize the usage of exotic matter to construct a wormhole, and it is known as ‘Cut and Paste’ technique,  in which the exotic matter is concentrated at the wormhole throat. In the  ‘Cut and Paste’ technique, the wormholes can be theoretically constructed by cutting and pasting two different  manifolds to obtain geodesically complete new manifold with a throat placed in the joining shell \cite{Poisson1995}. Using the Darmois-Israel formalism \cite{Israel1966,Darmois1927}, one can determine the surface stresses of the exotic matter located in thin shell placed at the joining surfaces. This new construction of wormhole is known as thin shell wormhole and is extremely useful to perform the stability analysis  for the dynamic case. Several authors have used this surgical technique (i.e., the Cut and Paste technique) to construct thin wormholes. Poisson and Visser \cite{Poisson1995} have analyzed the stability of a thin wormhole constructed by joining two Schwarzschild spacetimes. Eiroa and Simeone \cite{Eiroa2005} have constructed the wormholes by applying the cut and paste technique to two metrics corresponding to a charged black hole solution in low energy bosonic string theory, with vanishing antisymmetric ﬁeld along with the Maxwell ﬁeld, whereas they have analyzed cylindrically symmetric thin wormhole geometry associated to gauge cosmic strings \cite{Eiroa2004}. Thibeault et al. \cite{Thibeault2006} have analysed the stability as well as the energy conditions of ﬁve dimensional spherically symmetric thin shell wormholes in Einstein–Maxwell theory  along with the Gauss Bonnet term.

Now, the Einstein-Gauss-Bonnet (EGB) gravity which is a  generalization of Einstein's general relativity, was originally introduced  by Lanczos \cite{Lanczos1938}, and it was rediscovered by David Lovelock \cite{Lovelock1971}. Amid the larger class of general higher-curvature theories, this theory has some special characteristics having higher derivative terms. Nevertheless, the field equations in this theory are of second-order 
like that of  general theory of relativity. In this theory the Gauss-Bonnet term is present in the low energy effective action of heterotic string theory \cite{Callan1986,Witten1985,Gross1987}, and can be considered as the dimensionally extended version of the four-dimensional Euler density. Also, this term appears in six-dimensional Calabi-Yau compactifications of \textit{M}-theory \cite{Guica2006}. The EGB gravity enables the researchers to study how higher curvature corrections to black hole physics significantly change the qualitative features which are available due to black holes under the background of general relativity.

The Gauss-Bonnet term naturally appears as the second significant term in heterotic string effective action. Also, the noncommutative geometry appears  from the study of open string theories. One can show that the gravitational wave signal GW150914 which has been recently detected by LIGO and Virgo collaborations \cite{Abbott2016},  can be used to place a bound on the scale of quantum fuzziness of noncommutative space-time \cite{Kobakhidze2016}. In a paper \cite{Kobakhidze2016}, author has shown that the leading noncommutative correction to the phase of the gravitational waves produced by a binary system appears at the second order of the post-Newtonian expansion . Also, in another paper \cite{Saha2015}, the plausibility of using quantum mechanical transitions that is induced by the combined effect of gravitational waves  and noncommutative  structure  to probe the spatial noncommutative nature  has been explored. Synder \cite{Snyder1947a,Snyder1947b} first introduced the noncommutative spacetime to study the divergences in relativistic quantum field theory. Noncommutative geometry also aims to place General Relativity and the Standard Model on the same footing in order to describe gravity, the electro-weak and strong forces as gravitational forces in a unified spacetime \cite{Stephan2013}.
Noncommutativity substitutes the  point-like structures by smeared objects \cite{Nicolini2006, Nicolini2010,Lobo2013,Rahaman2012} and is used to eliminate the divergences which normally appear in general theory of relativity. It is supposed to be  intrinsic property of spacetime and hence,  does not depend on any particular feature such as curvature.

Although, general relativity retains the usual commutative form, the noncommutative geometry leads to a smearing of matter distributions which can be considered as due to the intrinsic uncertainty embodied in the coordinate commutator as
\begin{equation}\label{nco}
	\left[x_{\alpha}, x_{\beta}\right] = i \theta_{\alpha\beta}, 
\end{equation}
where $\theta_{\alpha\beta} $ is an anti-symmetric matrix which determines the fundamental cell discretization of spacetime. The standard way of modelling the smearing effect is accomplished by using a Gaussian distribution of minimal length $\sqrt{\theta}$ due to possible uncertainty.
Hence, one can consider a mass $M$, described by a $\delta$-function 
distribution for a static, spherically symmetric, Gaussian-smeared matter source, in $D-$dimensions \cite{Spallucci2009,Rizzo2006}, as
\begin{equation}\label{eq2}
	\rho_{\theta}(r) = \frac{\mu }{(4 \pi \theta)^{(D-1)/2}} e^{{-r^2}/{(4\theta)}}
\end{equation} 
where $M$ is the total mass of the source, may be due to a diffused centralized object such as a wormhole \cite{Lobo2013} and the particle mass $ \mu $ is supposed to be  diffused throughout a region of linear size $ \sqrt{\theta} $. Here, $ \theta $ is the noncommutative parameter which can be taken to be of a Planck length. Thus, the above equation plays the role of a matter source and the mass is smeared around the region $ \sqrt{\theta} $ instead of situated at a particular point.

One can find a number of studies inspired by noncommutative geometry which are available in the literature either under the background of general relativity ~\cite{Garattini2009,Rahaman2015,Gladney2017} or under modified gravity theories~\cite{Sharif2013,Hassan2021, Samir2020} .The model of self-sustained traversable wormhole has been studied by R. Garattini and F.S.N. Lobo \cite{Garattini2009}, whereas Rahaman et. al \cite{Rahaman2014,Rahaman2015} explored  two new wormhole solutions inspired by noncommutative geometry taking two different energy density profile admitting conformal motion. P.K.F. Kuhfittig and V. D. Gladney \cite{Gladney2017} studied charged wormhole inspired by noncommutative geometry with low tidal forces. In another paper authors \cite{Sharif2013} have studied the womhole solutions inspired by noncommutative geometry in the framework of $f(T)$ gravity whereas Hassan et al. have  \cite{Hassan2021} studied the womhole solutions in teleparallel gravity  inspired by noncommutative geometry. In a work Samir et al.\cite{Samir2020} have investigated the existence of wormholes by introducing noncommutative geometry in terms of Gaussian and Lorentzian distributions in $f(R,G)$ gravity. The present paper searches for a solution of ﬁve-dimensional (5D) Einstein-Gauss-Bonnet equations in the presence of a static, spherically symmetric Gaussian mass distribution to ﬁnd a thin-shell wormhole solution inspired by the noncommutative geometry. The statring point is to solve the non-linear Einstein-Gauss-Bonnet equations for a static and spherically symmetric metric for a source as mention by Eq.(\ref{eq2}).

The paper is organized as follows: In Section 2 we ﬁnd ageneral solution which leads to the five dimensional (5D) spherically symmetric static Einstein-Gauss-Bonnet equations for the source as mentioned by Eq.(\ref{eq2}), which further allows us to compute some quantities, viz., the componets of stress-energy tensors around the thin shell of  a noncommutative inspired thin-shell wormhole in Section 3. In Section 4 we find the Equation of State (EOS) at the throat of the thin-shell wormhole. We investigate the motion test particle near the throat to find the nature wormhole in Section 5 whereas Section 6 deals with the total exotic matter contet at the shell. We analyse the linearized stability of our model in Section 7 and finally, in Section 8 we pass some concluding remarks.

\section{Black holes in non-commutative geometry  inspired Einstein-Gauss-Bonnet  gravity}\label{S:HLBH}.

The  action of  Einstein-Gauss-Bonnet gravity in D-dimension \cite{Deser1985,Wheeler1986} can be written as

\begin{equation}\label{Action}
	\mathcal{S}=\frac{1}{2\kappa}\int dx^{D}\sqrt{-g}\left[  \mathcal{L} +\alpha \mathcal{L}_{GB}
	\right] + \mathcal{S}_{m},
\end{equation}

$\mathcal{S}_{m}$ represents the action associated with matter and $\alpha$ is a
coupling constant with dimension of $(\mbox{length})^2$  which is positive in the hetoretic string theory. For the case of general theory of relativity   $ \mathcal{L} = R$, the Ricci scalar  and the second-order Gauss-Bonnet term $\mathcal{L}_{GB}$ is given by
\begin{equation}
	\mathcal{L}_{GB}=R_{\mu\nu\gamma\delta}R^{\mu \nu\gamma\delta}-4R_{\mu\nu}R^{\mu\nu}+R^{2}.
\end{equation}
Throughout the paper we assume that $8 \pi G=c=1$ and the variation of the above action with respect to the metric $g_{\mu\nu}$ gives the Einstein-Gauss-Bonnet equations \cite{Cai2004,Vaz2006,Ghosh2008} as

\begin{equation}\label{ee}
	G_{\mu\nu}^{E}+\alpha G_{\mu\nu}^{GB}=T_{\mu\nu},
\end{equation}

where $G_{\mu\nu}^{E}$ is the Einstein tensor, while $G_{\mu\nu}^{GB}$ is explicitly given
by \cite{Kastor2006}
\begin{eqnarray}
	G_{\mu\nu}^{GB} & = & 2\;\Big[ -R_{\mu\sigma\kappa\tau}R_{\quad\nu}^{\kappa
		\tau\sigma}-2R_{\mu\rho\nu\sigma}R^{\rho\sigma}-2R_{\mu\sigma}R_{\ \nu
	}^{\sigma} +RR_{\mu\nu}\Big] -\frac{1}{2}\mathcal{L} _{GB}g_{\mu\nu}. 
\end{eqnarray}
Here, one can note that the divergence of Einstein-Gauss-Bonnet tensor $G_{\mu \nu}^{GB}$ vanishes and the spherically symmetric vacuum solutions \cite{Deser1985,Wheeler1986} for this theory is given by 

\begin{equation}
	ds^2=  - f(r) dt^2+ \frac{dr^2}{ f(r)} + r^2 d\Omega_D^2
\end{equation}

where $ d\Omega_D^2 $ is the line element on the D unit sphere, i.e.,

\begin{equation}
	d\Omega_D^2 =  d{\theta}_1^2 + \sin^2{\theta}_1 d
	{\theta}_2^2+ ...... + \prod_{n=1}^{D-1}\sin^2
	{\theta}_n d{\theta}_D^2
\end{equation}

and volume of the D unit sphere is given by  

\begin{equation}
	\Omega_D =  2\frac{ \pi
		^\frac{D+1}{2}}{\Gamma(\frac{D+1}{2})}
\end{equation}

As we have mentioned earlier that noncommutativity eliminates point like structure in 
favor of smeared objects in the flat spacetime. The effect of smearing is mathematically 
implemented with a Gaussian distribution of minimal width $\sqrt{\theta}$
\cite{Nicolini2006}. Here, we find  5 dimensional (5D) black hole solution inspired by the noncommutative geometry in the Einstein-Gauss-Bonnet gravity. For the matter density $\rho$  mentioned before along with the condition $g_{00}= 1/g_{rr}$ two components of the diagonal stress-energy tensor reads
\begin{eqnarray}\label{cemt}
	T^0_0 &=& - T^r_r=  \rho_{\theta}(r) = \frac{\mu }{16 \pi^2 \theta^2} e^{{-r^2}/{(4\theta)}}. 
\end{eqnarray}
The Bianchi identity  $T^{ab};b=0$~\cite{Rizzo2006} reads 
\begin{equation}
	0= \partial_r T^r_r + \frac{1}{2} g^{00} \left[T^r_r-T^0_0\right] \partial_r g_{00} 
	+ \frac{1}{2} \sum g^{ii}  \left[T^r_r-T^i_i\right] \partial_r g_{ii} 
\end{equation}
and due to the condition $g^{ii} \partial_r g_{ii} = 2/r$, one can obtain the energy momentum tensor in the five dimensional (5D) spacetime as 
\begin{eqnarray}\label{EMT}
	T^t_t &=& T^r_r= \rho_{\theta}(r), 
	\nonumber \\
	T^{\theta}_{\theta} &=& T^{\phi}_{\phi} = T^{\psi}_{\psi} = \rho_{\theta}(r) + \frac{r}{3} 
	\partial_r \rho_{\theta}(r), 
\end{eqnarray}
and the energy-momentum tensor  is specified by Eq. (\ref{EMT}). 

Moreover, Eq.~(\ref{ee}) for the matter source as given by Eq. (\ref{EMT}), leads to a general solution  \cite{Spallucci2009}
\begin{equation} \label{sol:egb}
	f_{\pm}(r) = 1+\frac{r^{2}}{4{\alpha}}\left[1\pm \sqrt{1+\frac{8\alpha \mu}{ r^4 \pi} 
		\gamma\left(2, \frac{r^2}{4\theta}\right) }\right],
\end{equation}

where $\gamma\left(2 \ , r^2/4\theta\, \right)$ is the lower incomplete Gamma function which is defined as
\begin{equation}
	\gamma\left(2\ , r^2/4\theta\, \right)\equiv \int_0^{r^2/4\theta}\; du\; u\; e^{-u}.  
\end{equation}
To proceed further, one can  define the mass-energy $\mu(r)$ as
\begin{equation}\label{mass}
	\mu(r) = \frac{2  \mu}{\pi} \gamma\left(2\ , r^2/4\theta\, \right),
\end{equation}
and  the total mass-energy ($M$) measured by asymptotic observer \cite{Ansoldi2006} is given by
\begin{equation}\label{mass1}
	M= \lim_{r \rightarrow \infty} \mu(r) = \frac{2  \mu}{\pi} ,
\end{equation}
and the solution as given by Eq. (\ref{sol:egb}) reads
\begin{equation} \label{sol:egb1}
	f_{\pm}(r) = 1+\frac{r^{2}}{4{\alpha}}\left[1\pm \sqrt{1+\frac{8\alpha M}{r^{4}}
		\gamma\left(2, \frac{r^2}{4\theta}\right) }\right],
\end{equation}

Here, if one consider $M_{1}=\frac{M}{\sqrt{\theta}}$, $\alpha_{1}=\frac{\alpha}{\sqrt{\theta}}$ and $r_{1}=\frac{r}{\sqrt{\theta}}$, then the metric function takes the form

\begin{equation}
	f(r)=1+\frac{r_1^2 \left(1-\sqrt{\frac{8 \alpha _1 M_1-2 \alpha _1 M_1
				e^{-\frac{r_1^2}{4}}
				\left(r_1^2+4\right)}{r_1^4}+1}\right)}{4 \alpha _1}
\end{equation}

\begin{figure}[ht!]
	\centering
	\begin{minipage}[b]{1.2\textwidth}
		\includegraphics[width=.8\textwidth]{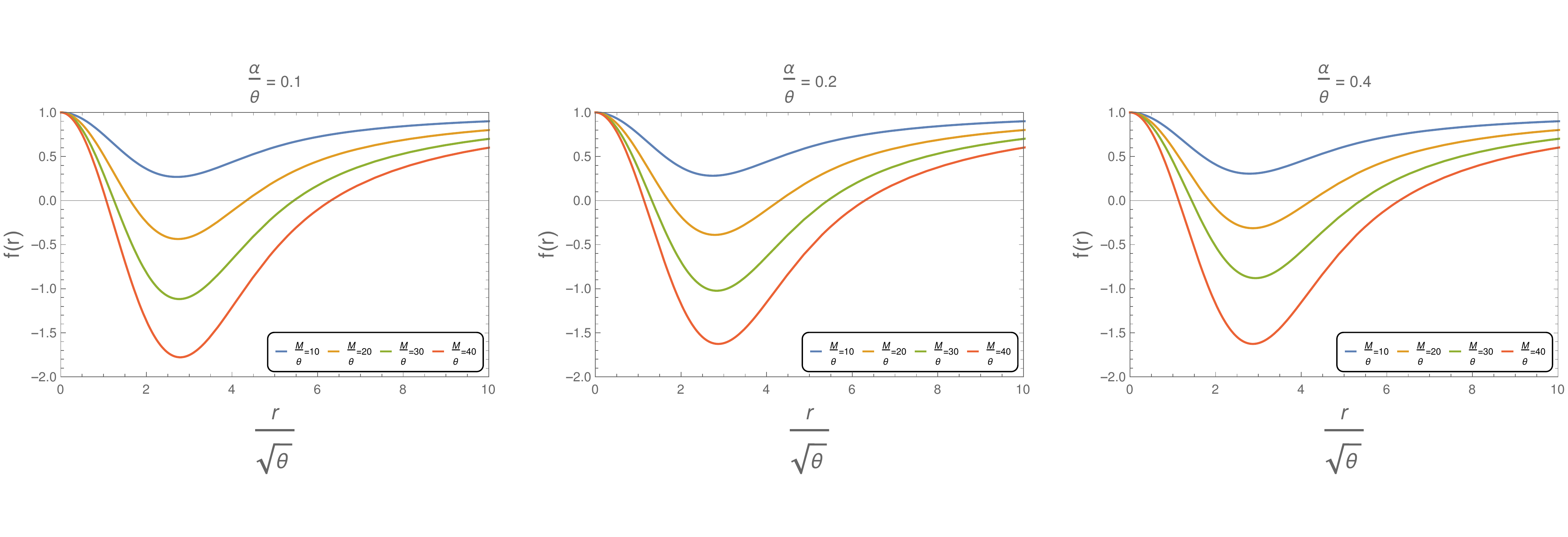}
	\end{minipage}\qquad
	\caption{The variation in the metric function \textbf{f(r)} with respect to \textbf{r} $(km)$ .}
	\label{fig:1}
\end{figure}

In Fig. (1) the function $f(r)$ has been plotted w.r.t. $r$ for different values of $\frac{\alpha}{\sqrt{\theta}}$ and from the figure it is clear that the  difference between the values of $f(r)$ at the inner horizon with that at outer horizon  shows an increasing profile with the increase in $\frac{M}{\sqrt{\theta}}$ value. There also exits a critical value for $\frac{M}{\sqrt{\theta}}$ to form a black hole and from the it is revealed that there is  no significant dependence on $\alpha$ at  two horizons.

\section{Thin-shell wormholes from black holes in non-commutative geometry  inspired Einstein-Gauss-Bonnet  gravity }\label{S:thin-shell}

The thin shell wormhole that we are to construct can be done by coping the solutions of two regular black holes and the region can be written as (removing from each of the copies)
\begin{equation}\label{4}
	\Omega^\pm \equiv\{ r \leq a \mid a > r_{h} \},
\end{equation}
where $a$ is a constant and $r_{h}$ is the event horizon of the black hole. As a result of the above operation, the timelike hypersurfaces can be expressed topologically,
\begin{equation}\label{5}
	d \Omega^\pm \equiv \{r=a\mid a>r_h\}
\end{equation}
The two timelike hypersurfaces located at $ d \Omega^{+} =d \Omega^{-} \equiv \ d \Omega $ along with two asymptotically flat regions that are connected by a wormhole whose throat is at  $ d \Omega $.\
The intrinsic metric upon $ d \Omega $ can be written as
\begin{equation}\label{6}
	ds^{2} =- d\tau^{2} + a^{2}(\tau) d \Omega^{2}_{D},
\end{equation}
where $ \tau $ is the proper time on the junction shell and $ a(\tau)$ is the radius of the throat.
Using Lanczos equation\cite{Israel1966,Darmois1927,lanczos1924,sen1924,perry1992,lake1996} the intrinsic surface strees energy tensor can be obtained as

\begin{equation}\label{7}
	S^{i}_{j} = -\frac{1}{8\pi} (k^{i}_{j} -\delta^{i}_{j}\ k^{l}_{l})
\end{equation}

where $k^\pm_{ij} = K^{+}_{ij} -K^{-}_{ij}$ is the discontinuity in the second fundamental form.
We can write the second fundamental\cite {rahaman2006,rahaman2009,usmani2010,rahaman2010,dias2010,rahaman2011} form as

\begin{equation}\label{8}
	K_{ij}^\pm =  - n_\nu [ \frac{\partial^2X_\nu}
	{\partial \xi^i\partial \xi^j } +
	\Gamma_{\alpha\beta}^{\nu^\pm} \frac{\partial X^\alpha}{\partial \xi^i}
	\frac{\partial X^\beta}{\partial \xi^j }]
\end{equation}

where $ n_\nu $ is the unit normal vector to $ d \Omega $ and $ \xi^i $ represents the intrinsic co-ordinates. The parametric equation at the hypersurface $ d \Omega $ is $ f(x^\mu(\xi^i)) =0 $.

Using the above equation, the normal vector can be expressed as

\begin{equation}\label{9}
	n_\mu^\pm =  \pm   | g^{\alpha\beta}\frac{\partial F}{\partial X^\alpha}
	\frac{\partial F}{\partial X^\beta} |^{-\frac{1}{2}} \frac{\partial F}{\partial X^\mu}
\end{equation}

with $ n^\mu n_\mu =+ 1 $.
$ k_{ij} $ can be written as $ k^{i}_{j} = diag  (k^\tau_\tau , k^{\theta_1}_{\theta_1}  k^{\theta_2}_{\theta_2} ,...., k^{\theta_D}_{\theta_D}) $, as a result of spherical symmetry.

Therefore, the surface-energy tensor has a form as $ S^{i}_{j} = diag (-\sigma ,p_{\theta_1},p_{\theta_2},.....,p_{\theta_D})$, where $\sigma $ is the surface energy density and $p$ is the surface pressure. Now from the Lanczos equation, we can get
\begin{equation}\label{10}
	\sigma = -\frac{D}{4\pi a}\sqrt{f(a)+ \dot{a}^2}
\end{equation}

\begin{equation}\label{11}
	p=p_{\theta_1}= p_{\theta_2}......=p_{\theta_D} =\frac{1}{8\pi} \frac{2 \ddot{a} +f^\prime(a)}{\sqrt{f(a)+ \dot{a}^2}} - \frac{D-1}{D}\sigma,
\end{equation}

where the dots and primes denote the differentiation with respect to $\tau$ and $a$ respectively.

The conservation equations can be written as
\begin{equation}\label{12}
	\frac {d}{d \tau} (\sigma a^D) + p \frac{d}{d \tau}( a^D)= 0
\end{equation}
or
\begin{equation}\label{13}
	\dot{\sigma} +  D\frac{\dot{a}}{a}( p + \sigma ) = 0,
\end{equation}
which are obeyed by $\sigma$ and $p$.
For the static configuration having a radius $a$, we take $\dot a =0$ and $\ddot a = 0$. All the calculations are done in five dimensions (5D), i.e., by putting $D=3$. Now  Eq.( \ref{10}) and Eq.(\ref{11}) reduces to
\begin{equation}\label{14}
	\sigma = -\frac{3}{4\pi a}\sqrt{f(a)}
\end{equation}

\begin{equation}
	\sigma(a)=-\frac{3 \sqrt{\frac{a^2 \left(1-\sqrt{\frac{8 \alpha  M
						\left(1-\frac{e^{-\frac{a^2}{4 \theta }} \left(a^2+4 \theta
							\right)}{4 \theta }\right)}{a^4}+1}\right)}{4 \alpha }+1}}{4
		\pi  a}
\end{equation}
Now, for  $\sigma_{1}=\sqrt{\theta} \sigma$ we have
\begin{equation}
	\sigma_{1}=-\frac{3 \sqrt{4-\frac{a^2 \left(\sqrt{\frac{a^4-2
						\left(a^2+4\right) e^{-\frac{a^2}{4}} \alpha  M+8 \alpha 
						M}{a^4}}-1\right)}{\alpha }}}{8 \pi  a}
\end{equation}
where $a_{1}=\frac{a}{\sqrt{\theta}}$
and
\begin{equation}\label{15}
	p =\frac{1}{8\pi} \frac{f^\prime(a)}{\sqrt{f(a)}} - \frac{2}{3}\sigma
\end{equation}

\begin{small}
	\begin{equation}
		p(a)=\frac{4 \alpha  M e^{-\frac{a^2}{4 \theta }} \left(4 \theta +a^2\right)+\theta  \left(-\left(16 \alpha  M+3 a^4 \right)+\left(3a^{2}+ 8 \alpha  a^2 \right) \sqrt{\frac{8
					\alpha  M-\frac{2 \alpha  M e^{-\frac{a^2}{4 \theta }} \left(4 \theta +a^2\right)}{\theta }+a^4}{a^4}}\right)}{8 \pi  \alpha  \theta 
			a^3\left( \sqrt{\frac{8 \alpha  M-\frac{2 \alpha  M e^{-\frac{a^2}{4 \theta }} \left(4 \theta +a^2\right)}{\theta }+a^4}{a^4}}\right)
			\sqrt{4-\frac{a^2 \left(\sqrt{\frac{8 \alpha  M-\frac{2 \alpha  M e^{-\frac{a^2}{4 \theta }} \left(4 \theta +a^2\right)}{\theta
							}+a^4}{a^4}}-1\right)}{\alpha }}}.
	\end{equation}
\end{small}
Similarly, for $p_{1}=\sqrt{\theta} p$ we have
\begin{eqnarray}
	p_{1}=\frac{4e^{-\frac{a_1^2}{4}} \alpha_1 M_1 (a^2_1+4)-(16\alpha_1 M_1+3a^4_1)+(3a^4_1+8\alpha_1 a^2_1)\sqrt{\frac{2 \alpha _1 M_1\left(4- e^{-\frac{a_1^2}{4}}
				\left(a_1^2+4\right)\right)}{a_1^4}+1}}{8 \pi 
		\alpha _1 a_1^3 \left(\sqrt{\frac{2 \alpha _1 M_1\left(4- e^{-\frac{a_1^2}{4}}
				\left(a_1^2+4\right)\right)}{a_1^4}+1}\right)
		\sqrt{\frac{4\alpha _1+a_1^2 \left(1-\sqrt{\frac{2 \alpha _1 M_1\left(4- e^{-\frac{a_1^2}{4}}
						\left(a_1^2+4\right)\right)}{a_1^4}+1}\right)}{\alpha _1}}}
\end{eqnarray}

where the function $f(a)$ is given by Eq.(5). The negative sign in the surface density itself implies the existence of exotic matter in the shell  which automatically violates the null energy condition as well as the weak energy condition.

\begin{figure}[ht!]
	\centering
	\begin{minipage}[b]{1.2\textwidth}
		\includegraphics[width=.8\textwidth]{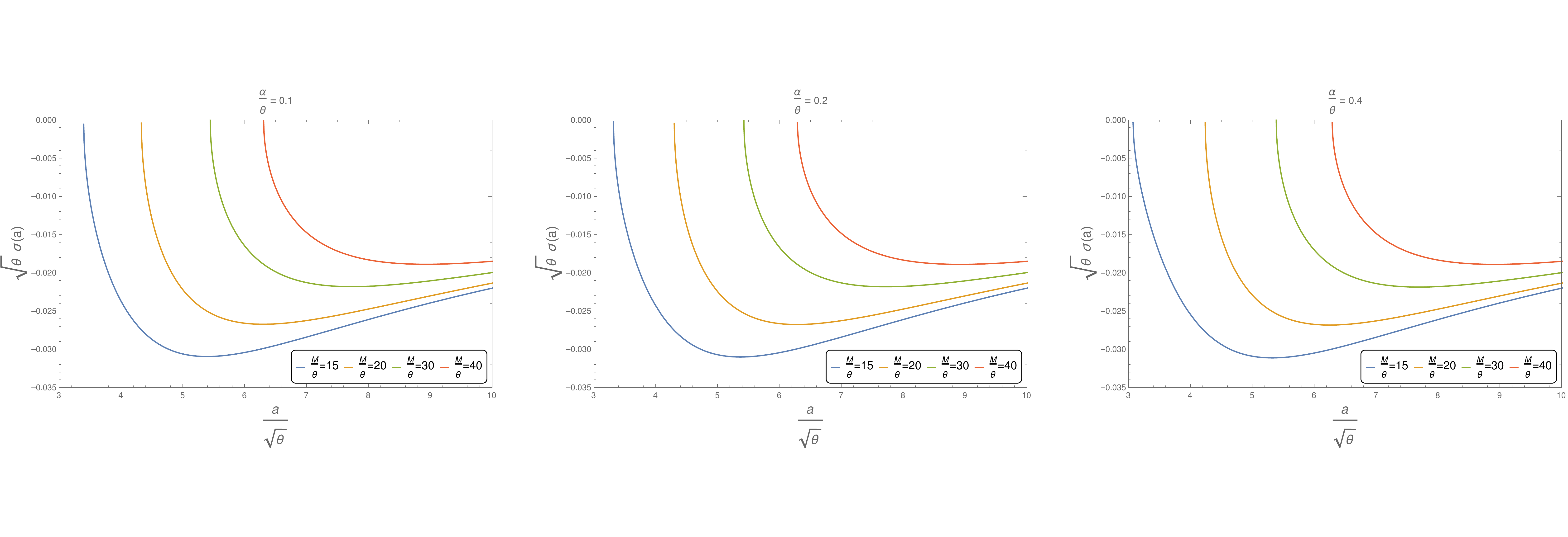}
	\end{minipage}\qquad
	\caption{The variation in {\boldmath$\sigma$} $(km^{-2})$ with respect to \textbf{a} $(km)$}
	\label{fig:2}
\end{figure}

In Fig. (2) we have plotted the surface energy density with respect to throat radius $a$ for different values of $\frac{M}{\theta}$ and from the figure it is revealed that as we are increasing the throat radius, the negative values of $\sigma$ increases rapidly and reaches a  maximum value and after that it slightly decreases and tends to a saturated value. Fig. (3) shows the variation of surface pressure with respect to throat radius $a$ and it is clear from the figure that as we increase the throat radius, the value of $p$ is decreasing rapidly but after a certain point it has no dependence on $\frac{M}{\theta}$ value. Fig. (4) represents the violation of the null energy condition which is a necessary for the construction of wormhole.

\begin{figure}[ht!]
	\centering
	\begin{minipage}[b]{\textwidth}
		\includegraphics[width=1\textwidth]{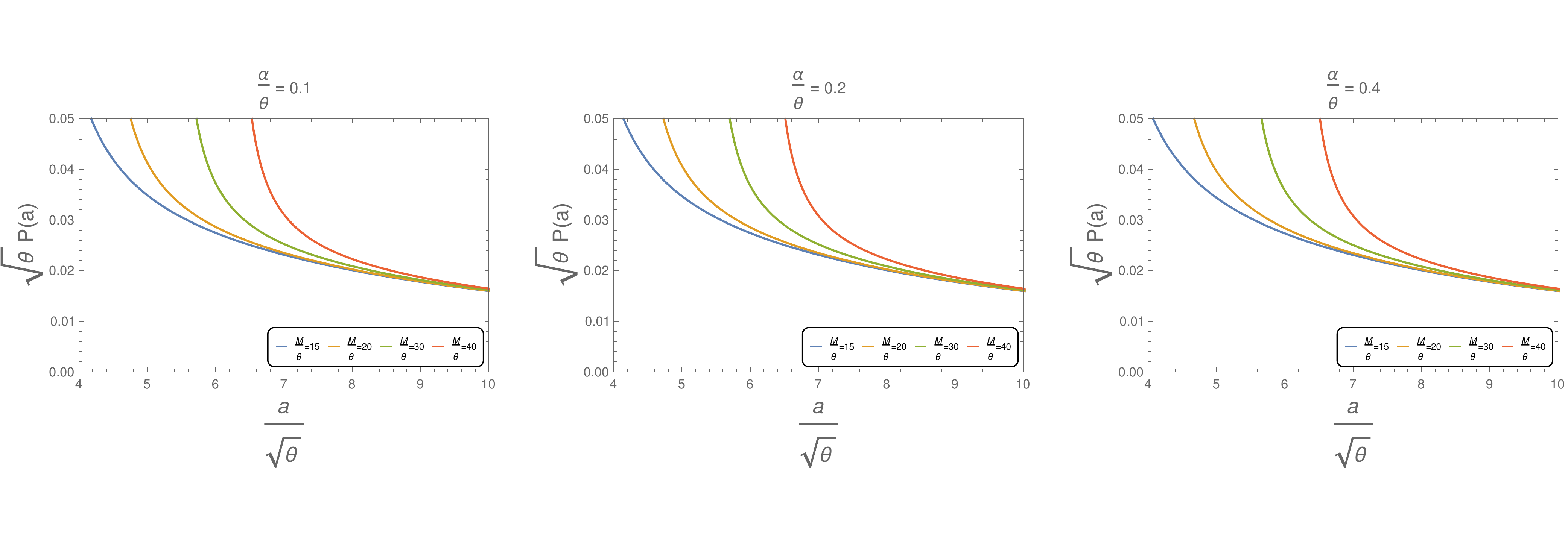}
	\end{minipage}\qquad
	\caption{Pressure \textbf{p} $(km^{-2})$ is plotted with respect to \textbf{a} $(km)$ }
	\label{fig:3}
\end{figure}

\begin{figure}[ht!]
	\centering
	\begin{minipage}[b]{1.2\textwidth}
		\includegraphics[width=.8\textwidth]{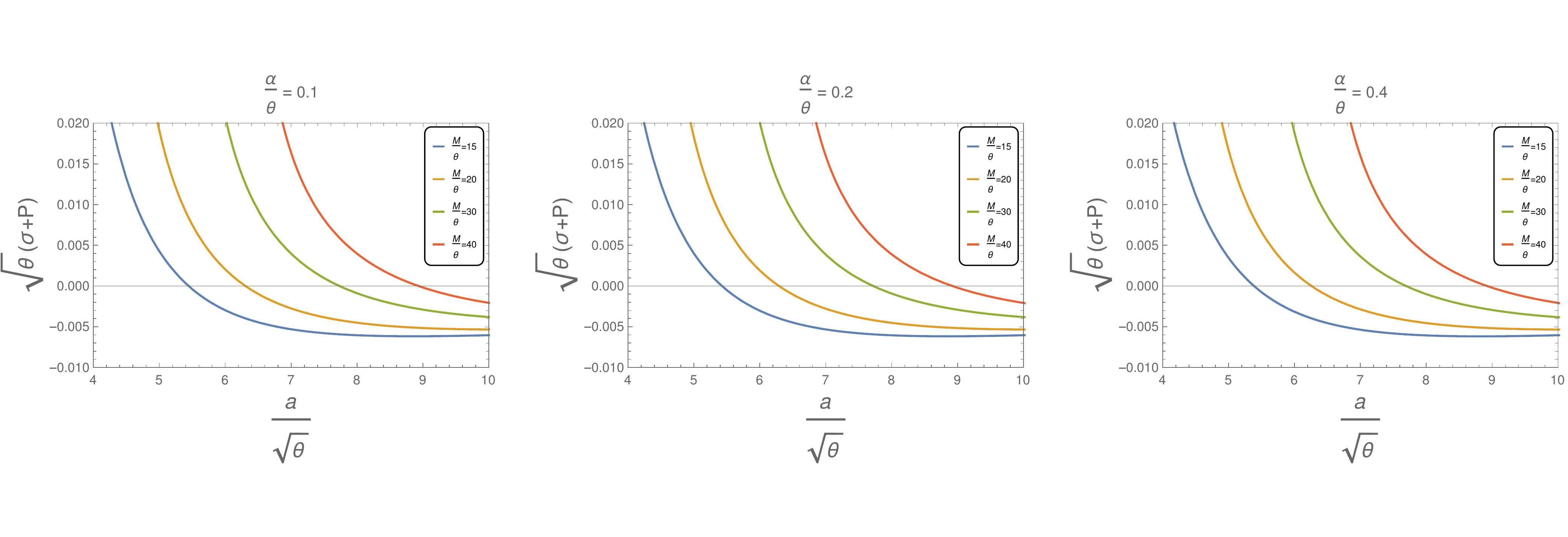}
	\end{minipage}\qquad
	\caption{{\boldmath$\sigma+p$} $(km^{-2})$ is plotted with respect to \textbf{a} $(km)$}
	\label{fig:4}
\end{figure}

\section{Equation of state}
The Equation of State (EOS)  is given by 
\begin{equation}
	p(a)= 	\omega~\sigma(a).
\end{equation}
\begin{figure*}[ht!]
	\centering
	\includegraphics[width=1.1\textwidth]{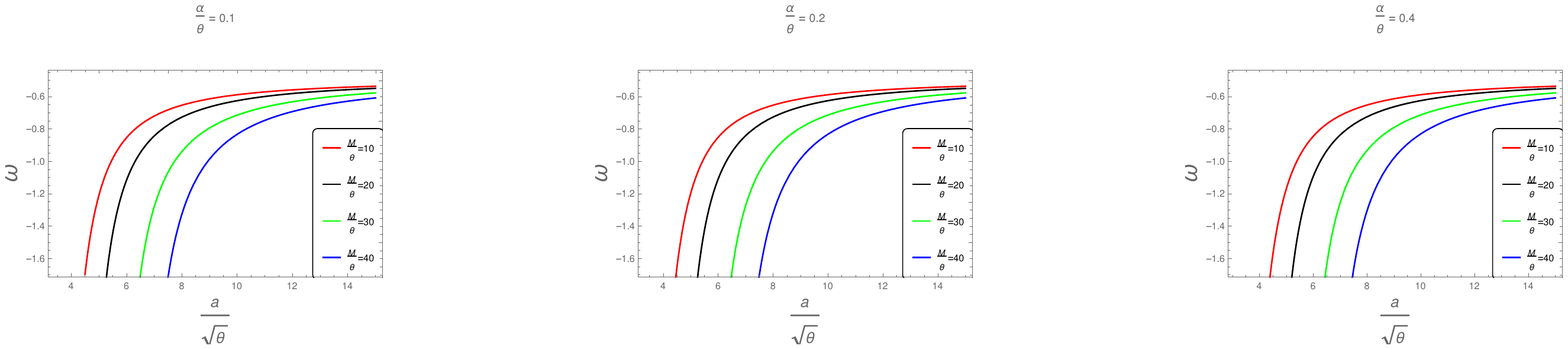}
	\caption{The variation of Equation of State (EOS) parameter with respect to \textbf{a} $(km)$}  
\end{figure*} 

Therefore, by virtue of Eqs. (31) and (34) the equation of state parameter $\omega$ can be explicitely written as 
\begin{equation}
	\omega= \frac{e^{-\frac{a^2}{4}} \left(4 \left(a^2+4\right) \alpha  M+e^{\frac{a^2}{4}} \left(3 a^4 \left(\sqrt{\frac{a^4-2 \left(a^2+4\right)
				e^{-\frac{a^2}{4}} \alpha  M+8 \alpha  M}{a^4}}-1\right)+8 \alpha  a^2 \sqrt{\frac{a^4-2 \left(a^2+4\right) e^{-\frac{a^2}{4}} \alpha 
				M+8 \alpha  M}{a^4}}-16 \alpha  M\right)\right)}{3 a^2 \sqrt{\frac{a^4-2 \left(a^2+4\right) e^{-\frac{a^2}{4}} \alpha  M+8 \alpha 
				M}{a^4}} \left(a^2 \left(\sqrt{\frac{a^4-2 \left(a^2+4\right) e^{-\frac{a^2}{4}} \alpha  M+8 \alpha  M}{a^4}}-1\right)-4 \alpha \right)}
\end{equation}

Fig. (5) shows the variation of equation of state parameter with respect to throat radius for different values of  $\frac{\alpha}{\theta}$ and it reveals that the negative values of $\omega$ decreases as we increase the throat radius. Also, there is no significant dependence on $\frac{\alpha}{\theta}$ values.

\section{The gravitational field}

To study the nature of the gravitational field of our wormhole we calculate the observer's four acceleration which is given by 

\begin{equation}\label{16}
	A^{\mu}= u^{\mu}_{;\nu} u^{\nu},
\end{equation}

where 

\begin{equation}\label{17}
	u^{\nu}= \frac{\partial x^\nu}{\partial\tau} = (\frac{1}{\sqrt{f(r)}},0,0,0,0).
\end{equation}

Hence, the only non-zero component can be written as 

\begin{equation}\label{18}
	A^{r} = \frac{f'(r)}{2} 
\end{equation}

or

\begin{equation}
	A^{r}=\frac{e^{-\frac{a^2}{4 \theta }}a \left(4 \theta ^2  e^{\frac{a^2}{4 \theta
		}} \left(\sqrt{\frac{8 \alpha  M
				\left(1-\frac{e^{-\frac{a^2}{4 \theta }} \left(4 \theta +a^2\right)}{4
					\theta }\right)}{a^4}+1}-1\right)-\alpha  M
		\right)}{16 \alpha  \theta ^2 \sqrt{\frac{2 \alpha  M
				\left(4-\frac{e^{-\frac{a^2}{4 \theta }} \left(4 \theta
					+a^2\right)}{\theta }\right)}{a^4}+1}}
\end{equation}
Now $A^{r}_{1}=\sqrt{\theta}A^r$, therefore,

\begin{equation}
	A^{r}_{1}=\frac{e^{-\frac{a_1^2}{4}} a_1 \left(4
		e^{\frac{a_1^2}{4}}\left( \sqrt{\frac{2 \alpha _1 M_1
				\left(4-e^{-\frac{a_1^2}{4}} \left(a_1^2+4\right)\right)}{a_1^4}+1}-1\right)-\alpha _1 M_1\right)}{16 \alpha _1 \sqrt{\frac{2 \alpha _1 M_1
				\left(4-e^{-\frac{a_1^2}{4}} \left(a_1^2+4\right)\right)}{a_1^4}+1}}
\end{equation}

For a  test particle moving towards the  radial direction from rest the geodesic equation can be written as
\begin{equation}\label{19}
	\frac{d{^2} r}{d\tau^2} = -\Gamma^{r}_{tt} \left(\frac{dt}{d\tau}\right)^{2} = -a^{r}
\end{equation}

The nature of the wormhole is attractive if $A^{r}>0$ and repulsive  if $A^{r}<0$ depending on the parameters of equation. We have plotted $A^{r}$ with respect to throat radius in Fig. (5) and it shows that the model of wormhole constructed is attractive in nature under the noncommutative geometry inspired Einstein-Gauss-Bonnet gravity. Also, the attractive nature is decreasing with the increase of throat radius.

\begin{figure}[ht!]
	\centering
	\begin{minipage}[b]{1.2\textwidth}
		\includegraphics[width=.8\textwidth]{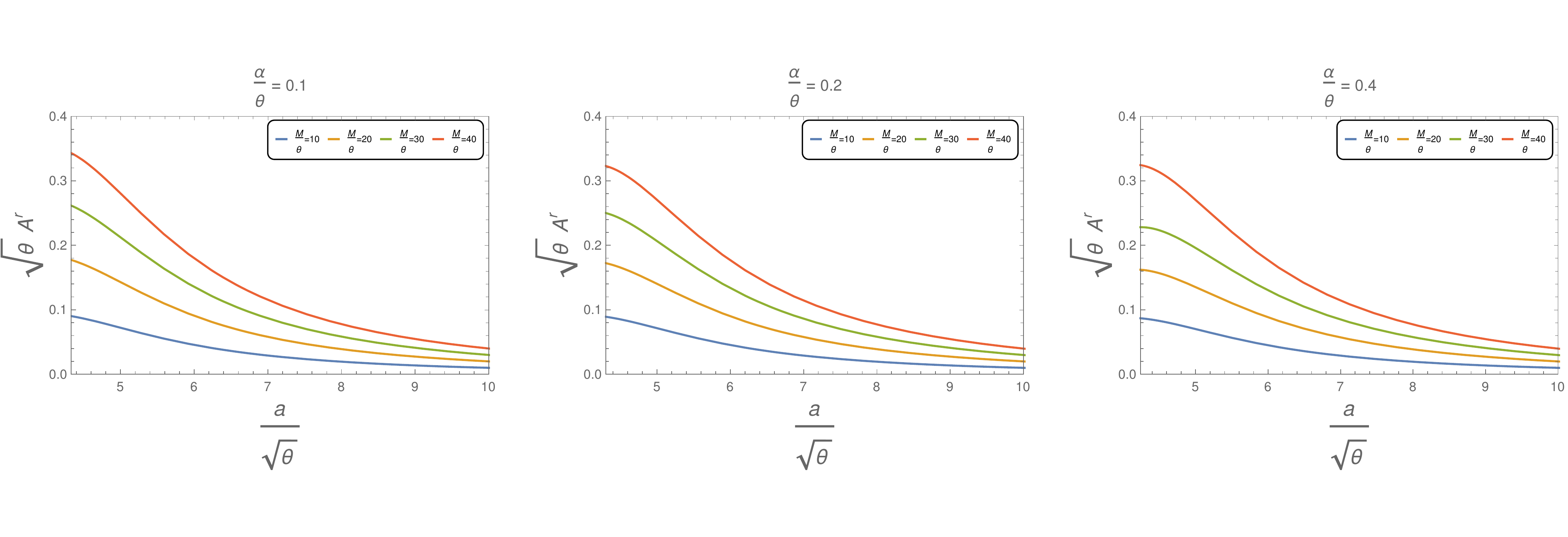}
	\end{minipage}\qquad
	\caption{Accelaration  \textbf{$a^{r}$} $(km)^{-1}$ of the test particle with respect to \textbf{a} $(km)$ }
	\label{fig:5}
\end{figure}

\section{The total amount of exotic matter}

We can determine the total amount of exotic matter in a thin shell wormhole by using the integral 
\begin{equation}\label{20}
	\Omega_{\sigma}=\int [\sigma + p]
	\sqrt{-g}d^{D+1}x
\end{equation}

A new radial co-ordinate $R=\pm (r-a)$ has been introduced by Eiroa and Simeone \cite{Eiroa2005} as
\begin{equation}\label{21}
	\Omega_{\sigma} =  \int_0^{2\pi} \int_0^{\pi}..... \int_0^{\pi}\int_{-\infty}^\infty [\rho + p]
	\sqrt{-g} dR d{\theta_1} d{\theta_2}.....d{\theta_D}
\end{equation}

Since, the radius of the shell is constant it does not exert any radial pressure, so that we have $ \rho =
\delta(R) \sigma(a)$ and one can write
\begin{equation}\label{22}
	\Omega_{\sigma}  = \int_0^{2\pi} \int_0^{\pi}..... \int_0^{\pi}\rho \sqrt{-g}
	]|_{r=a} d{\theta_1} d{\theta_2}.....d{\theta_D} =2a^D \sigma(a)\frac{\pi^{\frac{D+1}{2}}}{\Gamma(\frac{D+1}{2})}
\end{equation}
\begin{figure*}[ht!]
	\centering
	\begin{minipage}[b]{1.2\textwidth}
		\includegraphics[width=.8\textwidth]{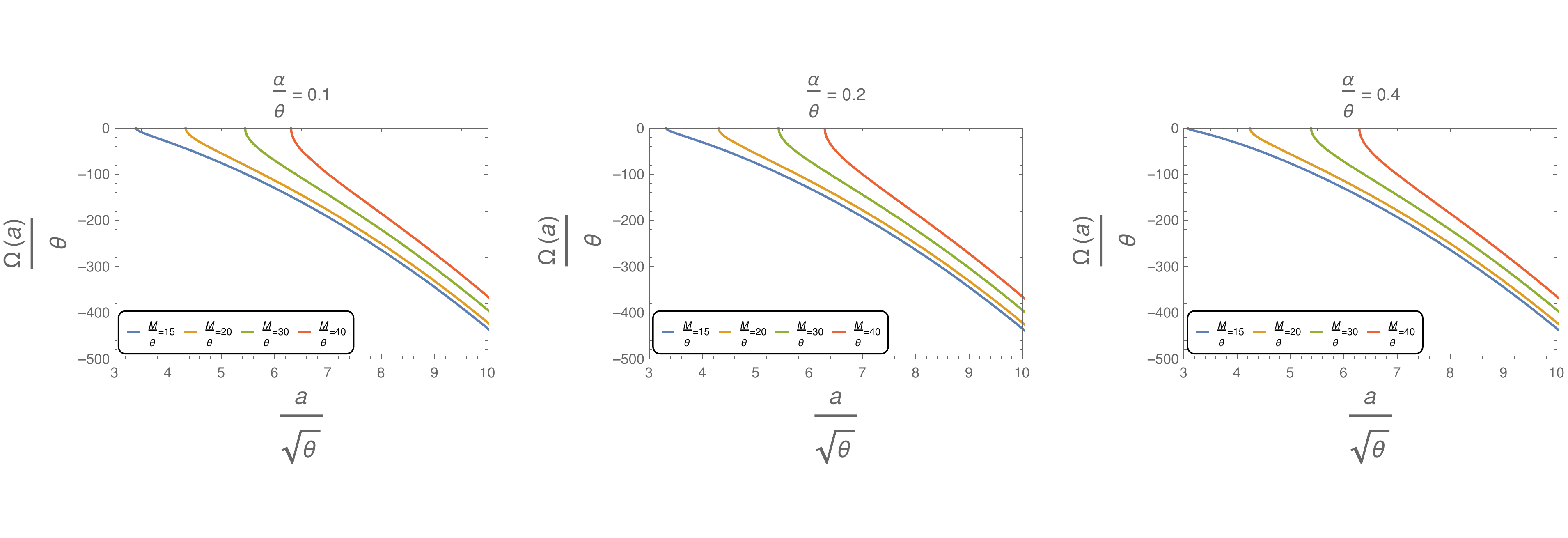}
	\end{minipage}\qquad
	\caption{The variation of the total amount of exotic matter $\Omega$ at the shell with respect to \textbf{a} $(km)$. }
	\label{fig:6}
\end{figure*}

\begin{equation}
	\Omega=-\frac{3}{2} \pi  a^2 \sqrt{\frac{a^2 \left(1-\sqrt{\frac{8 \alpha  M \left(1-\frac{e^{-\frac{a^2}{4 \theta }} \left(4 \theta +a^2\right)}{4
						\theta }\right)}{a^4}+1}\right)}{4 \alpha }+1}
\end{equation}
Now $\Omega_{1}=\frac{\Omega}{\theta}$ where
\begin{equation}
	\Omega_{1}=-\frac{3}{4} \pi  a_1^2 \sqrt{4-\frac{a_1^2
			\left(\sqrt{\frac{8 \alpha _1 M_1-2 \alpha _1 M_1
					e^{-\frac{a_1^2}{4}}
					\left(a_1^2+4\right)+a_1^4}{a_1^4}}-1\right)}{\alpha _1}}
\end{equation}

Fig.(7) shows the variation of total exotic matter content with respect to throat radius and from figure we can assert that as we increase the value of throat radius the amount of exotic matter at the shell  increases rapidly. So, to minimize the amount of exotic matter content we have to consider the throat of the wormhole near the event horizon.

\begin{figure}[ht!]
	\centering
	\begin{minipage}[b]{0.9\textwidth}
		\includegraphics[width=.8\textwidth]{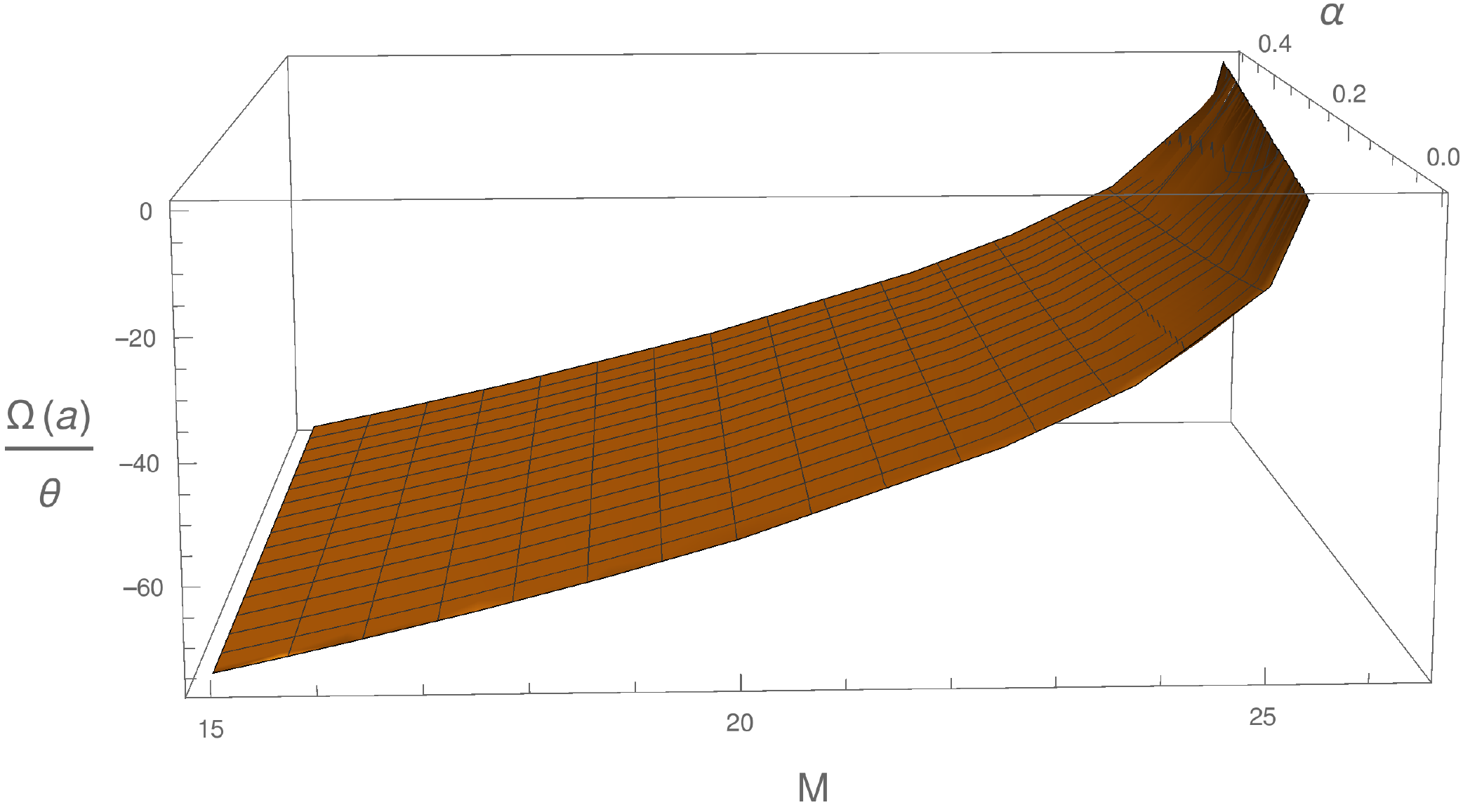}
	\end{minipage}\qquad
	\caption{Three dimensional plot of total exotic matter $\Omega$ with respect to total mass $M$ as well as $\alpha$.}
	\label{fig:7}
\end{figure}

\section{Linearized stability}
To obtain the stability of the thin shell wormhole one has to perform a small perturbation around a static solution at $a=a_{0}$. One can rearrange the   Eq. (25) to obtain the equation of
motion at the thin shell region as  

\begin{equation}  \dot{a}^2 + V(a)= 0  \end{equation}

where  the potential can be defined  as

\begin{equation}
	V(a) =  f(a) - \frac{16\pi^2 a^2\sigma^2(a)}{D^2}.
\end{equation}

Now, expanding the potential $V(a)$  into a Taylor series around $a_{0}$ upto second order we have

\begin{equation}\label{26}
	V =  V(a_0) + V^\prime(a_0) ( a - a_0) + \frac{1}{2} V^{\prime\prime}(a_0)
	( a - a_0)^2 + \mathcal{O}[( a - a_0)^3]
\end{equation}

where the prime $(')$ represents the derivative with respect to $a$.

\begin{figure}[ht!]
	\centering
	\begin{minipage}[b]{1.25\textwidth}
		\includegraphics[width=.8\textwidth]{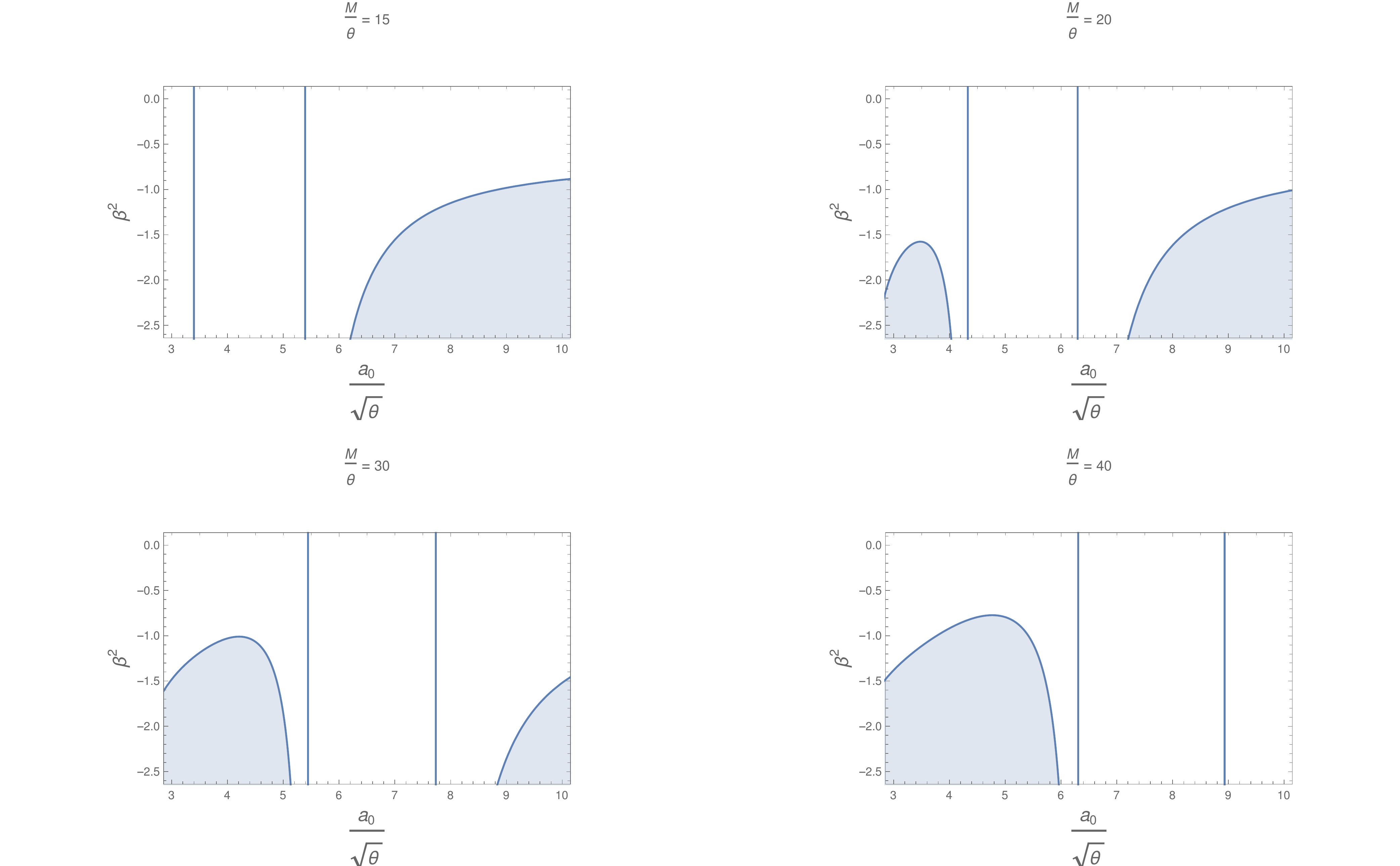}
	\end{minipage}\qquad
	\caption{Plots of {\boldmath$\beta^{2}$} as function of {\boldmath{$a_{0}$}} for $\alpha=0.1$}
	\label{fig:8}
\end{figure}

As we are linearizing around a static solution at $ a = a_0 $, we have $ V(a_0) = 0 $ and $ V^\prime(a_0)= 0 $ and the stable equilibrium configurations correspond to the condition $V^{\prime\prime}(a_0)> 0 $. Here, one can introduce a the parameter $\beta^{2}(\sigma)$ as the speed of sound which is given by
\begin{equation}\label{27}
	\beta^{2}(\sigma) = \frac{\partial p}{\partial \sigma}=\frac{p^{\prime}}{\sigma^{\prime}} 
\end{equation}

The first and second order derivatives of $V(a)$ is found to be
\begin{equation}\label{30}
	V^{\prime}(a)= f^{\prime}(a)-\frac{32 \pi^{2} a \sigma}{D^2}\left[\sigma-D(\sigma+p)\right]
\end{equation}
and
\begin{equation}\label{31}
	V^{\prime \prime}(a)=f^{\prime \prime}(a)-\frac{32\pi^2}{D^2}\left[\{(D-1)\sigma +D p\}^2-D \sigma(\sigma+p)(D-1+ D \beta^2)\right]
\end{equation}

Here,  $V(a_{0})=0$ and $V^{\prime}(a_{0})=0$ and for a stable wormhole solution one must have $V^{\prime\prime}(a_{0})>0$, i.e.,  

\begin{equation}\label{33}
	\beta^{2} <\frac{D^2 f''\left(a_0\right)-32 \pi ^2 \left(D^2 (p+\sigma ) (p+2 \sigma )-3 D \sigma  (p+\sigma )+\sigma ^2\right)}{32 \pi ^2 D^2 \sigma 
		(p+\sigma )}
\end{equation}

\begin{figure}[ht!]
	\centering
	\begin{minipage}[b]{1.25\textwidth}
		\includegraphics[width=.8\textwidth]{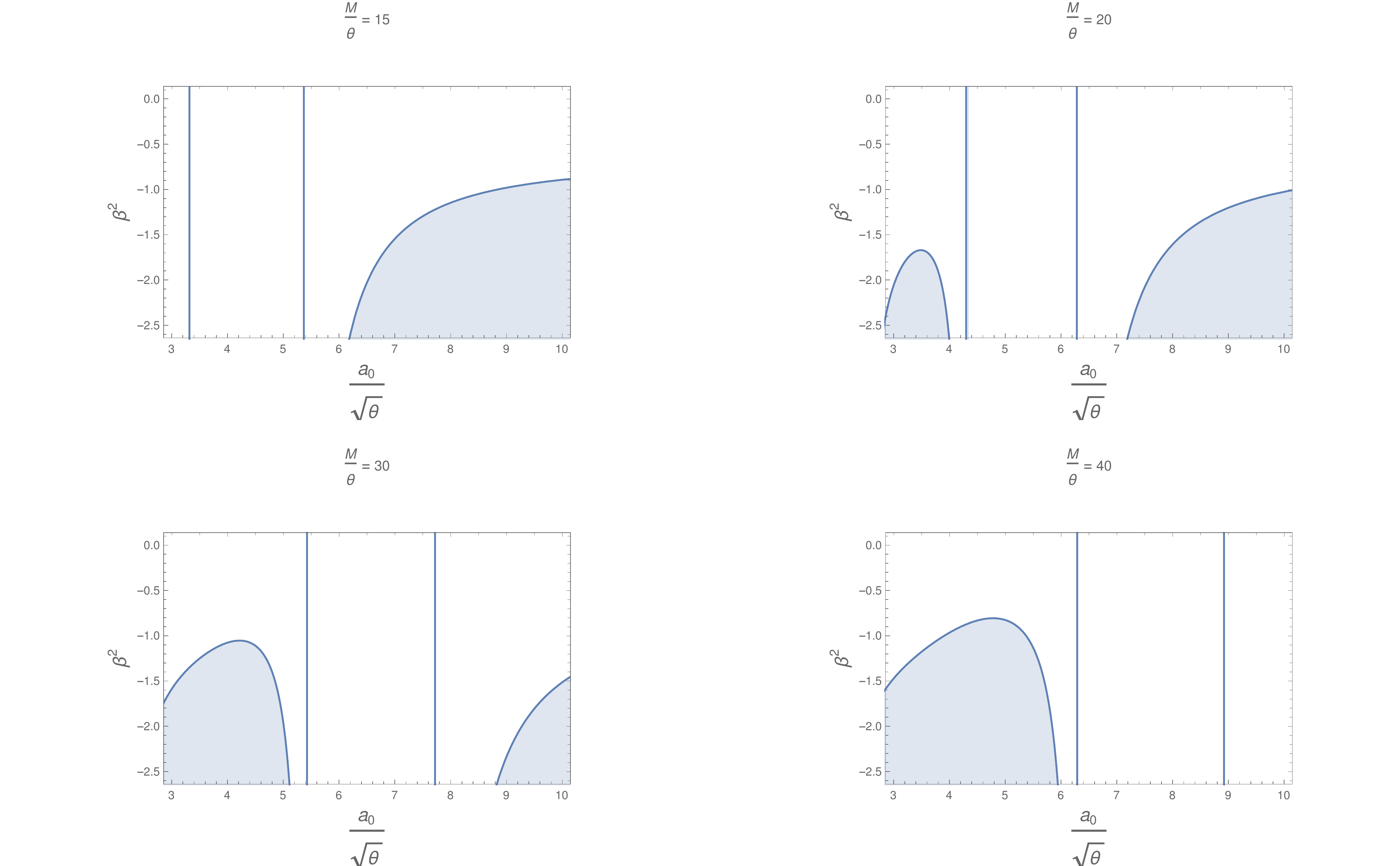}
	\end{minipage}\qquad
	\caption{Plots of {\boldmath$\beta^{2}$} as function of {\boldmath{$a_{0}$}} for $\alpha=0.2$}
	\label{fig:9}
\end{figure}

Using Eqs. (30) and (33) the above inequality takes the form
\begin{equation}\label{34}
	\beta^{2} < -\frac{a^2_0 (f_0^{\prime})^2-2 a_0 f \left(a_0 f_0''+(D-1) f_0'\right)+4 (D-1)
		f^2_0}{2 D f_0 \left(2 f_0-a_0 f_0'\right)},
\end{equation}

where $f_{0}=f(a_{0})$ , $f_{0}^{\prime}=f^{\prime}(a_{0})$ and $f^{\prime \prime}_{0}=f^{\prime \prime}(a_{0})$. Figs. (9) , (10), and (11) shows the stability regions of the wormhole solutions. From all these figures one can infer that as one increases the $\frac{M}{\theta}$ value the stability region is shifted from the higher value of throat radius to the lower value of throat radius $a$. Also, there is no significant change in the nature of the stability region as one varies the value of $\alpha$ which is evident from the figures.

\begin{figure}[ht!]
	\centering
	\begin{minipage}[b]{1.25\textwidth}
		\includegraphics[width=.8\textwidth]{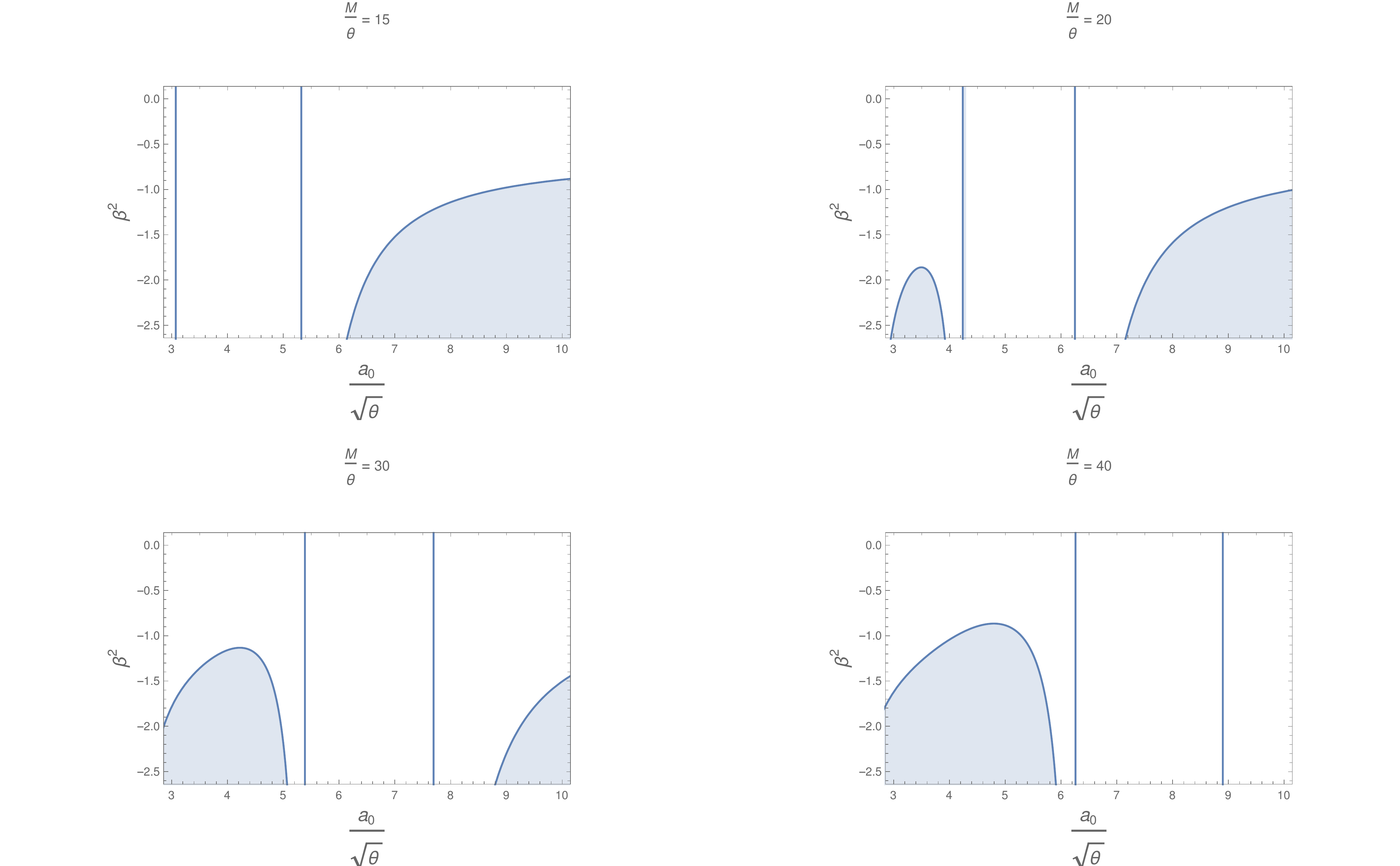}
	\end{minipage}\qquad
	\caption{Plots of {\boldmath$\beta^{2}$} as function of {\boldmath{$a_{0}$}}for $\alpha=0.4$}
	\label{fig:10}
\end{figure}

\section{Conclusion}

In this paper we construct a new thin-shell wormhole from black holes in non-commutative geometry inspired Einstein-Gauss-Bonnet gravity. For that 
we find out the non-commutative geometry inspired black hole solution in Einstein-Gauss-Bonnet gravity. Then, the `Cut and Paste' technique which was first introduced by M. Visser \cite{Visser1989},  is employed to construct the wormhole from the corresponding black hole spacetime assumed to be asymptotically flat in nature.

For the above mentioned structural form of a thin shell wormhole we have noted down several salient aspects of the solution as can be described below:\\

\indent(1) {\bf Pressure-density profile}: the surface energy density with respect to throat radius $a$ has been plotted in Fig. (2) for different values of $\frac{M}{\theta}$ and the figure shows that the negative values of $\sigma$ increases rapidly with the increase of throat radius and reaches a  maximum value and after that it slightly decreases and tends to attain a saturated value. Fig. (3) shows the variation of surface pressure with respect to throat radius $a$ and reveals that as we increase the throat radius, the value of $p$ is decreasing rapidly but after a certain point it has no dependence on $\frac{M}{\theta}$ value. Also, Fig. (4) represents the violation of the null energy condition which is a necessary for the construction of wormhole.\\

(2) {\bf Equation of state}: Fig. (5) which shows the variation of equation of state parameter with respect to throat radius for different values of  $\frac{\alpha}{\theta}$ reveals that the negative values of $\omega$ decreases as we increase the throat radius. Also, there is no significant dependence on $\frac{\alpha}{\theta}$ values.\\

(3) {\bf The gravitational field}: The acceleration of test particle $a^{r}$ has been plotted with respect to throat radius$a$ in Fig. (5) which shows that the model of wormhole constructed is attractive in nature under the noncommutative geometry inspired Einstein-Gauss-Bonnet gravity as  $a^{r}>0$ through out the entire range of $a$. Moreover, the attractive nature is decreasing with the increase of throat radius.\\

(4) {\bf The total amount of exotic matter}: Fig.(7) shows the variation of total exotic matter content with respect to throat radius $a$ and one can assert that as one increases the value of throat radius the amount of exotic matter at the shell  increases rapidly. So, to minimize the amount of exotic matter content one has to consider the throat of the wormhole near the event horizon.\\

(5) {\bf Linearized stability}:  In Figs. (9) , (10), and (11) we have depicted  the stability regions of the wormhole solutions for different values of  $\alpha$. From all these figures one can infer that as one increases the $\frac{M}{\theta}$ value the stability region is shifted from the higher value of throat radius to the lower value of throat radius $a$ and there is no significant change in the nature of the stability region as one varies the value of $\alpha$.\\

Finally, we can conclude that our model of thin-shell wormhole which is constructed from the `Cut and Paste' technique under the noncommutative geometry inspired Einstein-Gauss-Bonnet Gravity is found to be plausible with reference to the other model of thin-shell wormhole available in literature.

\begin{center}
	\textbf{Acknowledgment}
	
	NR is thankful to UGC MANF(MANF-2018-19-WES-96213) for providing financial support.FR \& MK would like to thank the authorities of the Inter-University Centre for Astronomy and Astrophysics, Pune, India for providing research facilities. 
\end{center}

\end{document}